# Runaway signals:
# Exaggerated displays of commitment may result from second-order signaling

Julien Lie-Panis[*a,b,c] and Jean-Louis Dessalles[b]

[a]*Institut Jean Nicod, Département d'études cognitives, Ecole normale supérieure, Université PSL, EHESS, CNRS, 75005 Paris, France*
[b]*LTCI, Télécom Paris, Institut Polytechnique de Paris, 91120 Palaiseau, France*

August 9, 2023

**Abstract**

To demonstrate their commitment, for instance during wartime, members of a group will sometimes all engage in the same ruinous display. Such uniform, high-cost signals are hard to reconcile with standard models of signaling. For signals to be stable, they should honestly inform their audience; yet, uniform signals are trivially uninformative. To explain this phenomenon, we design a simple model, which we call the signal runaway game. In this game, senders can express outrage at non-senders. Outrage functions as a *second-order signal*. By expressing outrage at non-senders, senders draw attention to their own signal, and benefit from its increased visibility. Using our model and a simulation, we show that outrage can stabilize uniform signals, and can lead signal costs to run away. Second-order signaling may explain why groups sometimes demand displays of commitment from all their members, and why these displays can entail extreme costs.

**Keywords**: signaling; commitment displays; ritual; game theory; outrage

[*]Corresponding author; Email: `jliep@protonmail.com`; ORCID: 0000-0001-7273-7704

# 1 Uniform investment in high-cost displays

Membership in human groups often involves ritual behaviors which appear arbitrary and wasteful to non-members, ranging from the embarrassment of hazing and the time constraints of religious practice to the emotional and physical scarring of certain rites or recruitment devices (Atran & Henrich, 2010; Cimino, 2011; Densley, 2012; Sosis et al., 2007; Whitehouse & Lanman, 2014). Drawing on honest signaling theory (Grafen, 1990; Spence, 1974; Veblen, 1899/1973; Zahavi, 1975), these behaviors have been explained as displays of prosocial commitment (Bulbulia & Sosis, 2011; Gambetta, 2009; Irons, 2001; Sosis, 2003).

Yet, some commitment displays seem uniform, in direct contradiction to the predictions of honest signaling theory. Displays of commitment are often binary. Individuals decide whether or not to participate in a rite, or whether or not to comply with a prescription. When in addition investment is universal, that is when all group members engage in the binary display, the resulting signal is uniform (at least in first approximation, see also: Barker et al., 2019). Uniform signals are trivially dishonest. In theory, they should not be stable.

In the next section, we introduce an explanation for uniform displays, based on understanding outrage as a second-order signal of commitment. To formally investigate our theory, we adapt Gintis, Smith and Bowles' (2001) multi-player model. When outrage is absent, signaling occurs only at an honest, non-uniform equilibrium, as shown in section 3. In section 4, we show that outrage can destabilize the honest signaling equilibrium, and lead to uniform signaling. In section 5, we introduce a simulation of our model, and show that outrage can also lead to high-cost displays, through a step-by-step runaway process[1]. We discuss the scope of our model in section 6.

# 2 Outrage as a second-order signal

Our aim here is to reconcile the existence of uniform displays with honest signal theory, based on a formal model. Mathematical signaling games have helped clarify the logic of a wide range of animal behaviors, pertaining for instance to mate choice (Grafen, 1990), cooperation (Leimar, 1997), aggression (Enquist, 1985), parent-offspring conflict (Godfray, 1991), and predator-prey interactions (Smith & Harper, 2003). In these models, interactions are most often dyadic, or involve one receiver and many signalers.

In contrast, the ritual behaviors we have mentioned occur in the context of an entire group. To model commitment displays, we adapt a model introduced by Gintis et al. (2001). This model is distinctive in applying to group interactions, involving many signalers and many receivers. Crucially, signalers compete for asymmetric affiliations (from here on: for followers). Optimal signaler behavior depends on the behavior of other signalers. In equilibrium, being the first to display is always beneficial, as one is able to attract many followers; in contrast, being the last to display is assumed to be net costly for individuals of low quality. As a result, a partial pooling equilibrium (Bergstrom & Lachmann, 1998) is obtained, in which individuals of lower quality opt out of the costly display entirely (Dessalles, 2014; Gintis et al., 2001).

This issue is exacerbated when the display is binary, as in Gintis et al.'s model (2001), and ours below. When individuals all invest in the same display, signaling is uniform, and therefore dishonest. Universal investment in a binary display should be doubly impossible. Not only should low quality signalers opt out of investing in a net costly display, but receivers should not pay attention to a entirely uninformative signal.

To explain universal investment in commitment displays, non-senders must face additional costs. We propose an endogenous source for those costs. In the type of group interaction we model, an individual's signal is susceptible to be observed by only a fraction of potential followers. Senders may be motivated to exploit non-senders, if this allows them to advertise their own signal beyond its direct observers. They may denounce or bully individuals who do not comply with a display to draw attention to their own compliance. As a result, non-senders face new costs. Universal investment could emerge out of a single motivation: advertising one's prosocial commitment, by any means necessary.

---

[1] We borrow the term runaway from Fisher (1915). The mechanism we have in mind is, however, entirely different from Fisher's. In our simulation, individuals gradually invest in higher levels of signaling due to social pressures—in order to attract partners, and avoid others' outrage.

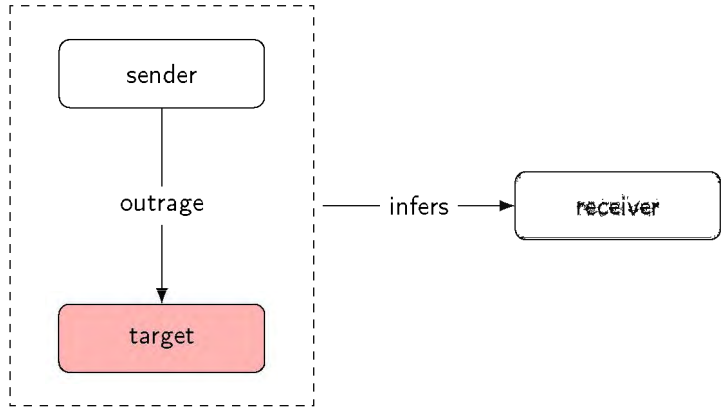

Figure 1: Outrage as a second-order signal. A sender can express outrage at a target who does not invest in the signal. When outrage is honest, receivers can infer that the sender has invested in the signal, even without having observed the sender's behavior directly. Outrage makes the sender's signal more visible. As a side-effect, the target is harmed.

More specifically, we argue that universal displays can be propped up by moral condemnation. Moral condemnation may take various forms, ranging from negative gossip to a dyadic partner about a third-party's immoral behavior to public expressions of collective outrage. It can entail a degree of reputational and/or material costs for its target. Here, we encapsulate these differing forms of moral condemnation and the associated costs for targets using the term outrage. Outrage can be a credible signal of moral behavior. To infer the moral quality of our partners, we sometimes use their propensity to verbally condemn a third-party's immoral behavior (Jordan et al., 2017). Conversely, to advertise our investment in desirable behavior, we sometimes express outrage against those who unambiguously display undesirable behavior (Jordan & Rand, 2019); or even against those whose morality is merely ambiguous (Jordan & Kteily, 2022).

In the context of commitment displays, outrage can be thought of as a *second-order signal* — a signal about (the absence of) a signal (Figure 1). We may for instance draw attention to those who secretly eat during a fast, and whose transgression may have otherwise gone unnoticed. In doing so, we not only broadcast our own investment, but we also indirectly increase others' incentive to display, thus laying the groundwork for universal, and even uniform, signaling.

# 3 Baseline model

## 3.1 A multi-player model of commitment displays with uncertain observation

To study commitment displays, we adapt the model introduced by Gintis et al. (2001). We consider a large group of individuals, who are characterized by a continuous quality $q$. We normalize minimum and maximum quality to 0 and 1 respectively: each individual's quality is drawn according to a continuous probability density function, whose support is $[0, 1]$. Individuals only observe their own quality. For mathematical convenience, the group is considered to be infinite in size.

Individuals alternate between two roles, that of signaler and receiver. Play occurs in three stages.

1. *Signaling stage.* Here, signalers decide whether to pay a cost $c_1(q)$ to send a signal, that is participate in a binary display of commitment. (The only other option is not to send.) Sending the signal is cheaper for high quality individuals: $c_1$ is a strictly decreasing continuous function of individual quality $q$, which takes positive values. In the present context, individuals of higher quality can be thought of as individuals who are more committed to the group and/or its moral values, and whose commitment translates into an increased ability or willingness to invest in the display—e.g. because the display will cause them to "burn bridges" with other groups, to which they are relatively uncommitted (Brusse, 2020).

2. *Observation stage.* Here, receivers do two things. First, they decide whether to pay a small positive cost $\nu > 0$ to monitor the signal. Second, receivers who paid the cost of monitoring observe the action chosen by each individual signaler in the previous stage, i.e. whether the signaler opted to send or not send. The probability of observation is $p_1$ ($0 < p_1 < 1$). Since the population is infinite, they observe the behavior of a fraction $p_1$ of signalers. As long as sending occurs with positive probability, monitoring receivers each observe at least one sender (not necessarily the same one). Receivers who did not pay the cost of monitoring do not observe behavior in the signaling stage.

3. *Social interaction stage.* Here, receivers decide whether to follow one signaler, that is to affiliate to one individual from the group. Signalers gain positive payoff $s > 0$ for each receiver who decides to follow them. Receivers derive payoff $f(q')$ from following a signaler of quality $q'$, and null payoff from opting not to follow anyone. Following is on average beneficial, and high quality individuals are more desirable social partners: we assume $\mathbf{E}(f) > 0$, and that $f$ is a strictly increasing continuous function of the followee's quality $q'$. Following low quality individuals may or may not be detrimental (depending on the sign of $f(0)$).

Signalers may decide to send the signal depending on their quality. Receivers may decide to: not monitor and not follow anyone; not monitor and follow any individual, i.e. follow an individual chosen at random; monitor and follow a sender, i.e. follow an individual chosen at random among those signalers they observed sending the signal; or monitor and follow a non-sender. A pure strategy profile specifies: (i) when in the signaler role, whether to send or not send given own quality $q$, and (ii) when in the receiver role, which one of the above four strategies to play. We do not consider mixed strategies, in which individuals behave probabilistically.

## 3.2 Honest signaling equilibrium

There are two evolutionarily stable strategy profiles (ESS; Maynard Smith & Price, 1973). First, the strategy profile in which: (i) signalers never send, and (ii) receivers do not monitor, and follow an individual at random. This trivial strategy profile is always a strict Nash equilibrium, and therefore always an ESS. Indeed, deviation to sending the signal is costly for all signalers, whatever their quality. Deviation to monitoring is costly for receivers, as is deviation to not following—because following an individual at random is beneficial ($\mathbf{E}(f) > 0$). We ignore this ESS from here on.

The second ESS is obtained by considering a family of strategy profiles. For any threshold quality $\theta \in (0, 1)$, we define the honest signaling strategy profile for $\theta$ as the strategy profile whereby: (i) signalers send when their quality $q$ verifies $q > \theta$, and do not send when $q < \theta$, and (ii) receivers monitor, and follow a sender. We note this strategy profile $\mathrm{HS}(\theta)$. We do not consider signaler strategy given $q = \theta$; since quality is continuously distributed, this occurs with null probability. When individuals play according to such a strategy profile, we define $\pi(\theta) \equiv \mathbf{P}(q > \theta) \in (0, 1)$ the probability that a signaler is of relatively high quality $q > \theta$, and sends.

We show that $\mathrm{HS}(\theta)$ is an ESS if and only if:

$$\nu < \mathbf{E}(f(q) \mid q > \theta) - \mathbf{E}(f) \tag{3.1}$$

$$c_1(\theta) = \frac{s}{\pi(\theta)} \tag{3.2}$$

Below, we outline the main steps allowing us to derive both conditions. We show that (3.2) at best defines a unique value of $\theta$, and therefore a single strategy profile to consider. The full demonstration is detailed in the Supplementary Information.

## 3.3 Uniform signaling is unstable

Condition (3.1) is obtained by considering the case of receivers. When signalers play according to $\mathrm{HS}(\theta)$, receivers pay the cost of monitoring, and follow an individual of relatively high quality $q > \theta$. On average, they gain: $\mathbf{E}(f(q) \mid q > \theta) - \nu$. In contrast, a rare mutant who opts not to monitor and follow any individual can expect to gain $\mathbf{E}(f)$. We deduce the proposed condition by comparing both payoffs.

For signaling to be evolutionarily stable, the relative benefit of conditioning affiliation on the signal must outweigh the cost of monitoring. This is a relatively weak condition. Since the cost of monitoring $\nu$ is small, condition (3.1) may be satisfied even when discrimination by the signal is weak (low positive threshold $\theta$), and even when partner quality is weakly associated to payoffs (small derivative $f'$).

In equilibrium, the signal is honest. When they observe the signal, receivers can infer the sender is of relatively high quality, above a certain positive threshold $\theta > 0$. In contrast, uniform signaling ($\theta = 0$) is always uninformative, and can therefore never be evolutionarily stable. If all signalers send, receivers learn nothing from the signal, and mutants who do not monitor can invade.

## 3.4 Existence of a signaling equilibrium

Condition (3.2) is obtained by considering the case of signalers. When receivers play according to HS($\theta$), signalers compete to attract followers by sending the signal. Each receiver observes a fraction $p_1 \times \pi(\theta)$ of senders, and chooses one to follow.

A signaler of quality $q$ who sends the signal pays cost $c_1(q)$ in the signaling stage. In the social interaction stage, that signaler is individually observed by each receiver with probability $p_1$. Each time the signaler is observed by a receiver, she is chosen with probability $\frac{1}{p_1 \times \pi(\theta)}$ (since the receiver chooses one individual at random among all observed senders), in which case she gains $s$. On average, she gains: $p_1 \times \frac{1}{p_1 \times \pi(\theta)} \times s$. The signaler's expected payoff is then: $-c(q) + \frac{s}{\pi(\theta)}$.

Since not sending is free, a rare mutant who deviates from HS($\theta$) by not sending given relatively high quality (resp. sending given relatively low quality) earns less than the resident when the above expression is positive for $q > \theta$ (resp. negative for $q < \theta$). We deduce that, for HS($\theta$) to be an ESS, the above expression must be null for $q = \theta$; this yields condition (3.2). In equilibrium, the signal is net beneficial for high quality Signalers, and net costly for low quality Signalers.

Condition (3.2) is an equation in $\theta$, with at best one solution. When $\theta$ varies from 0 to 1, $c_1(\theta)$ strictly decreases from $c_1(0)$, and $\frac{s}{\pi(\theta)}$ strictly increases to infinity from $\frac{s}{1} = s$ (because the distribution's support is the entire interval $[0, 1]$). Following the intermediate value theorem we obtain a unique solution $\theta \in (0, 1)$ if and only if:

$$c_1(0) > s \tag{3.3}$$

A signaling equilibrium, defined for the unique value of $\theta \in (0, 1)$ which solves equation (3.2), can only exist when the cost of sending for individuals of minimal quality ($q = 0$) is prohibitively high. Conversely, condition (3.3) guarantees the existence of a signaling ESS, so long as monitoring is sufficiently cheap, as per condition (3.1) (we detail the demonstration in the Supplementary Information).

# 4 The signal runaway game

## 4.1 Adding outrage to the baseline model

The signal runaway game occurs when we introduce outrage into the previous model. We view outrage as a *second-order signal*. Outrage refers to the commitment display (the first-order signal), by referring to a target's lack of signaling. Its function is to draw attention to the fact that the outraged individual did send the signal. Outraged senders increase everyone's incentive to send, and may destabilize the honest signaling equilibrium studied above. We modify the game in the following manner.

1. *Signaling stage.* Signalers decide whether to invest in costly signaling, as before, as well as whether to pay a cost $c_2 > 0$ to express outrage.

2. *Observation stage.* By expressing outrage, individuals draw attention to their signaling behavior. Signalers who paid the cost of second-order signaling $c_2$ are observed with increased probability $p_2 > p_1$ ($p_2 < 1$). In our model, onlookers can only observe whether an individual sent the signal—with probability $p_1$ or $p_2$, depending on whether the individual paid to express outrage.

Outraged signalers observe the signal, and select a target. Each individual's signaling behavior is thus observed by receivers who pay the cost of monitoring, as well as signalers who pay the cost of outrage. We assume outraged signalers select a target among all individuals they observe not sending the signal. Since the population is infinite, outraged signalers find a target in all situations but uniform signaling.

3. *Social interaction stage.* Targets of outrage are harmed. Signalers lose $h > 0$ for each individual who expresses outrage against them. As before, receivers can follow signalers.

A pure strategy for the signaler now specifies whether or not to send, and, if opting to send, whether to express outrage or not, depending on own quality $q$. For every $q$, there are three possibilities: send and express outrage, send and do not express outrage, and do neither. Receiver strategies are unchanged.

## 4.2 A note on our assumptions

Note that we do not consider the hypocritical strategy, whereby a signaler of quality $q$ does not send the signal, yet pays the cost of second-order signaling. Due to the simplified manner in which we model observation, this strategy is dominated. Receivers can only condition on a individual's observed signal, and not on whether the individual expresses outrage. Hypocritical signalers pay $c_2$ to draw attention to their lack of signaling, which is never beneficial in our model because it does not allow them to attract more followers.

Our model is intended to show the consequences of outrage functioning as a second-order signal, that is the consequences of onlookers using outrage to infer compliance (Jordan et al., 2017), for exogenous reasons. Nevertheless, we come back to the issue of hypocrisy in section 4.7, by extending our model.

The cost of outrage is fixed, and thought of as small. It is intended to reflect the risk of retaliation by the target (and, technically, the cost of monitoring, since outraged senders need to find a target). Note that targets are always non-senders. Unjustified punishment, which can damage one's reputation (Barclay, 2006), does not arise in our model. Instead, outraged individuals target non-senders, and earn a form of hard-coded reputational benefit, by increasing their chances of being followed when receivers value the signal itself.

Lastly, note that the cost of being outraged $h$ is exogenous. We view $h$ as encapsulating a variety of reputational and/or material costs which are suffered in contexts exterior to the model, ranging from the cost of being the subject of another indiviudal's negative gossip to the cost of constituting a legitimate target for collective punishment. In our simplified model, there can be no endogenous costs: targets of outrage can neither lose future followers nor attract more outrage in the future because all followees and targets of outrage are selected simultaneously, in the observation stage (besides, targets of outrage are non-senders who already do not attract any followers). The simulation presented in section 5 implements richer dynamics, allowing for such costs. It clarifies that even when allowing for such endogeneous costs, the exogenous cost of being outraged $h$ must be positive for uniform signaling to occur (see Figure 3).

## 4.3 Honest signaling with outrage equilibrium

For any threshold quality $\theta \in (0, 1)$, we define the honest signaling with outrage strategy profile HSO($\theta$) as the strategy profile whereby: (i) signalers send and express outrage when their quality verifies $q > \theta$, and neither send nor express outrage when $q < \theta$, and (ii) receivers monitor, and follow a sender.

We show that HSO($\theta$) is an ESS if and only if:

$$\nu < \mathbf{E}(f(q) \mid q > \theta) - \mathbf{E}(f) \tag{4.1}$$

$$c_1(\theta) + c_2 = \frac{s}{\pi(\theta)} + \frac{\pi(\theta)h}{1 - \pi(\theta)} \tag{4.2}$$

$$c_2 < \frac{p_2 - p_1}{p_2} \times \frac{s}{\pi(\theta)} \tag{4.3}$$

The proof is analogous to the one before. Receiver strategy and trade-offs are unchanged, yielding condition (4.1), which is identical to (3.1). Each receiver observes a fraction $p_2 \times \pi(\theta)$ of senders, and chooses one to follow. An analogous calculation to the one before shows that a signaler of quality $q$ who sends and expresses outrage gains on average payoff: $-c_1(q) - c_2 + p_2 \times \frac{s}{p_2 \pi(\theta)} = -c_1(q) - c_2 + \frac{s}{\pi(\theta)}$.

Non-senders now face the cost of being potential targets of others' outrage. Each outraged signaler observes a fraction $p_1 \times (1 - \pi(\theta))$ of non-senders, and selects one as target. Non-senders face an outraged signaler with probability $\pi(\theta)$, and are observed by that individual with probability $p_1$. They can now expect a negative payoff, equal to: $p_1 \times \frac{\pi(\theta)(-h)}{p_1(1-\pi(\theta))} = -\frac{\pi(\theta)h}{1-\pi(\theta)}$. We obtain condition (4.2) by comparing to the payoff above. When $\theta$ verifies this condition, signalers of quality $q = \theta$ are indifferent between sending both the signal and the second-order signal, and sending neither. Deviation to sending neither signal given $q > \theta$ is then detrimental, as is deviation to sending both signals given $q < \theta$.

Finally, condition (4.3) is obtained by considering rare mutants who deviate to sending but not expressing outrage given $q > \theta$. Such an individual saves on the cost of second-order signaling $c_2$, but is observed with probability $p_1 < p_2$, earning only $p_1 \times \frac{s}{p_2\pi(\theta)}$ on average in terms of followers. Comparing to the payoff of an outraged sender, we obtain the proposed condition.

## 4.4 Sufficient condition for the evolution of outrage

Since $\frac{1}{\pi(\theta)} \geq 1$, we deduce a sufficient condition for (4.3), valid whatever the value of $\theta \in (0, 1)$:

$$c_2 < \frac{p_2 - p_1}{p_2} s \qquad (4.4)$$

We show that outrage can be expected to invade any honest signaling equilibrium under the same condition (see Supplementary Information). Outrage evolves when the cost of second-order signaling is small relative to the benefit of making one's signal more visible to followers. Under condition (4.4), we do not need to consider the send and do not express outrage strategy, which is dominated in any honest signaling equilibrium.

## 4.5 Outrage can destabilize the honest signaling equilibrium

Outrage perturbs the signaling equilibrium. Senders now compete to attract followers *and* evade others' outrage. Technically outrage could lead to less signaling—when the cost of expressing outrage is larger than the expected cost of being targeted by the outrage of others ($c_2 > \frac{\pi(\theta)h}{1-\pi(\theta)}$). Since $c_2$ is considered to be small, outrage will most often push more individuals to send the signal.

There are two possible outcomes, represented in Figure 2. First, when harm $h$ is low, outrage introduces a small perturbation, and we retain a separating equilibrium. Second, when the consequences of being the subject of others' outrage are dire, outrage introduces a larger perturbation—and may completely destabilize the honest signaling equilibrium. We show that, whatever the value of $\theta \in (0, 1)$, HSO($\theta$) is not an ESS if:

$$c_1(0) + c_2 < s + 2\sqrt{hs} \qquad (4.5)$$

The above condition is obtained by considering condition (4.2). Multiplying by $\pi(\theta)(1 - \pi(\theta))$, we obtain equivalently:

$$(c_1(\theta) + c_2 + h)\pi(\theta)^2 - (c_1(\theta) + c_2 + s)\pi(\theta) + s = 0$$

We recognize a second-order equation in $\pi(\theta)$, whose discriminant is:

$$\Delta = (c_1(\theta) + c_2 + s)^2 - 4(c_1(\theta) + c_2 + h)s = (c_1(\theta) + c_2 - s)^2 - 4sh$$

HSO($\theta$) cannot be an ESS when there is no solution $\theta \in (0, 1)$ to equation (4.2). A sufficient condition for that to occur is $\Delta < 0$. Since $c_1(\theta)$ increases when $\theta$ decreases, and since we necessarily have $c_1(0) + c_2 > c_1(0) > s$ (otherwise there is no signaling equilibrium to start from following condition (3.3)), we deduce that the squared term is positive when $\theta$ is sufficiently small. We can then take the squared root, and deduce a sufficient condition by replacing $\theta$ with 0; we obtain the proposed condition (4.5).

## 4.6 Uniform signaling can be stable when outrage harms ambiguous targets

Our main result is therefore negative: if outrage is sufficiently cheap to express, as per condition (4.4), and being the target of others' outrage is sufficiently costly, as per condition (4.5), then outrage invades, and fully destabilizes any honest signaling equilibrium. Under such conditions, there can be no signaling ESS. Uniform signaling remains impossible here, because the function of outrage is merely to attract more followers, and receivers stop monitoring the signal when it is uniform.

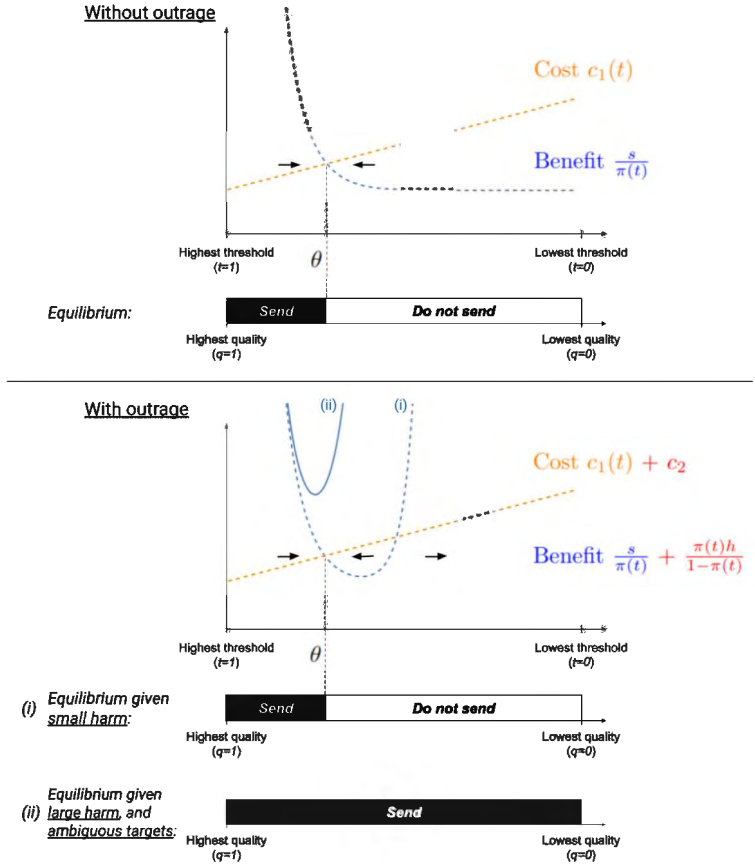

Figure 2: Effect of outrage on the signaling equilibrium. **Top:** In the absence of outrage, senders compete to attract followers. An honest signaling equilibrium is established at the threshold quality $\theta$ which equalizes cost (orange) and benefit of competing against a fraction $\pi(\theta)$ of senders to attract followers (blue). Note that we represent cost and benefit as a function of the potential threshold $t$, for $t$ varying between 1 and 0 (inverted x-axis). To the left of the graph, when $t$ is large, few high quality individuals ($q > t$) send. Going towards the right of the graph, as $t$ decreases, more and more lower quality individuals join in sending the signal. **Bottom:** Outrage increases the incentive to signal; senders compete to attract followers *and* evade others' outrage. Note that when $t$ tends towards 0, the benefit of evading others' outrage tends towards infinity since individuals then risk becoming the groups' moral punching bag; this is why the blue curve now takes on a U-shape. **(i)** When harm $h$ is low, we obtain another honest signaling equilibrium. The threshold quality $\theta$ is obtained at the first intersection of the orange and blue curves. To the left, when $t$ is just above $\theta$, too few high quality individuals send, and individuals whose quality is just below $t$ benefit from joining in. To the right, when $t$ is just below $\theta$, there are too many senders, and those of quality just above $t$ benefit from opting out. In contrast, the other intersection point is repellent. **(ii)** When harm is high, there is no honest signaling equilibrium. When in addition outrage can be directed at ambiguous targets, we obtain uniform signaling. For the purpose of illustration, we assume a linear cost function $c_1(q) = c_1(0) + q(c_1(1) - c_1(0))$, and that quality is normally distributed around $\bar{q} = 0.25$, with standard deviation 0.1. Other parameters: $c_1(0) = 3$, $c_1(1) = 1$, $s = 1$, $c_2 = 0.1$. In condition (i), we take $h = 0.01$; in condition (ii), we take $h = 0.1$.

Uniform signaling can however be made possible by extending the target selection mechanism. When all individuals signal, there are no non-senders to target. In our model, for technical reasons, this does not prevent signalers from investing in second-order signaling (because the model occurs in separate stages for simplicity, and we need outraged signalers' visibility to increase before the observation stage). We may instead assume that when individuals do not find non-senders, they use more ambiguous targets instead, in order to express outrage. Although outrage at ambiguous targets is less justified, and therefore riskier (Barclay, 2006), in some contexts it is used to attract reputational benefits (Jordan & Kteily, 2022).

We modify our model, by having outraged senders select as target: (1) a non-sender whom they observe, or, if they do not observe any non-sender (because signaling is uniform) (2) a signaler whose behavior they do not observe. Second-order signaling now serves two functions. Senders who are not observed not only miss out on possible followers but also risk being targeted by others' outrage.

We define the uniform signaling with outrage strategy profile USO as the strategy profile whereby: (i) senders signal and express outrage, whatever their quality, and (ii) receivers do not monitor, and follow an individual at random.

When individuals play according to USO, each signaler chooses an ambiguous target from the $1 - p_2$ percent of individuals that they do not observe. A signaler of quality $q$ pays both costs of signaling, and is a potential target of outrage for another individual with probability $(1 - p_2)$. That signaler earns average payoff: $-c_1(q) - c_2 - (1 - p_2) \times \frac{h}{1-p_2} = -c_1(q) - c_2 - h$.

Since the population is infinite, deviation to not sending is immediately detrimental: any signaler who attempts to save on the cost of sending risks becoming the group's moral punching bag, by constituting a preferential, unambiguous target for others' outrage. In addition, a rare mutant who deviates to not expressing outrage saves on cost $c_2$ but is unobserved, and therefore targeted, with increased probability $(1 - p_1) > (1 - p_2)$. On average, that mutant earns payoff: $-c_1(q) - (1 - p_1) \times \frac{h}{1-p_2}$. Comparing with the payoff of a resident, we deduce that USO is an ESS if and only if:

$$c_2 < \frac{(p_2 - p_1)h}{1 - p_2} \tag{4.6}$$

## 4.7 Outrage is honest when it targets hypocrites first

We can further extend the target selection mechanism to cater for hypocrites. Hypocrites are oft reviled, and judged more severely than individuals who admit to engaging in immoral behavior (Jordan et al., 2017). In the context of our model, hypocrites could constitute preferential targets for outrage.

Let us assume that hypocrites are preferential targets of outrage, that is that outraged individuals select as target: (0) an observed hypocrite, i.e. an individual whom they observe expressing outrage but not sending the signal, and, if no hypocrites are observed, (1) an observed non-sender, as before (and possibly, (2) an ambiguous target if they do not observe any non-sender).

In the model up until now, receivers cannot condition on others' outrage behavior, precluding any social benefits for hypocrites. Instead, let us assume the most favorable case for hypocrisy, that is that receivers follow at random among all individuals they observe either sending the signal, or expressing outrage—so long as they do not also observe a hypocrite not sending the signal. We assume that the probability that outrage is observed is $p' = \frac{p_2 - p_1}{1 - p_1}$ (such that $p_2 = p_1 + p' - p_1 p'$ is the probability that an outraged sender is observed sending either signal).

Under such conditions, HSO($\theta$) is immune to hypocrisy for every $\theta \in (0, 1)$. Since the population is infinite, deviation to not sending and expressing outrage is immediately detrimental, because this entails becoming a preferential target for a positive fraction of the population, and losing infinite payoff.

In addition, we can derive a sufficient condition given any positive fraction $\pi_H$ of hypocrites. Let us assume that, for a certain threshold $\theta \in (0, 1)$, signalers whose quality exceeds $\theta$ send and express outrage; and signalers whose quality is under $\theta$ never send, and express outrage with probability $\frac{\pi_H}{1-\pi(\theta)}$. We assume that receivers play as described above. This situation is analogous to the HSO($\theta$) strategy profile, with a total fraction $\pi_H > 0$ of hypocrites.

In such a situation, receivers each observe a total of $p_2\pi(\theta) + p'(1 - p_1)\pi_H$ potential followees, and chose one to follow; while outraged individuals each observe a fraction $p'p_1\pi_H$ of hypocrites, and chose one to target. On average, hypocrites earn: $-c_2 + p'(1-p_1)\frac{s}{p_2\pi(\theta)+p'(1-p_1)\pi_H} - p'p_1\frac{h}{p'p_1\pi_H}$. Since non-hypocritical non-senders earn null payoff (hypocrites concentrate outrage), a sufficient condition is obtained when, for hypocrites, the cost of facing other's outrage exceeds the benefit of attracting followers, that is (using $p'(1-p_1) = (p_2 - p_1)$) when:

$$\frac{h}{\pi_H} > \frac{s}{\pi_H + \frac{p_2}{p_2-p_1}\pi(\theta)}$$

The expression on the right is always smaller than $\frac{s}{\pi_H + 0}$. We deduce that, when senders express outrage and outrage is directed at hypocrites first, a sufficient condition for outrage to be honest is that the cost of being outraged exceed the benefit of attracting a follower, i.e. that:

$$h > s \tag{4.7}$$

# 5 Simulation

## 5.1 Outrage enables uniform signaling

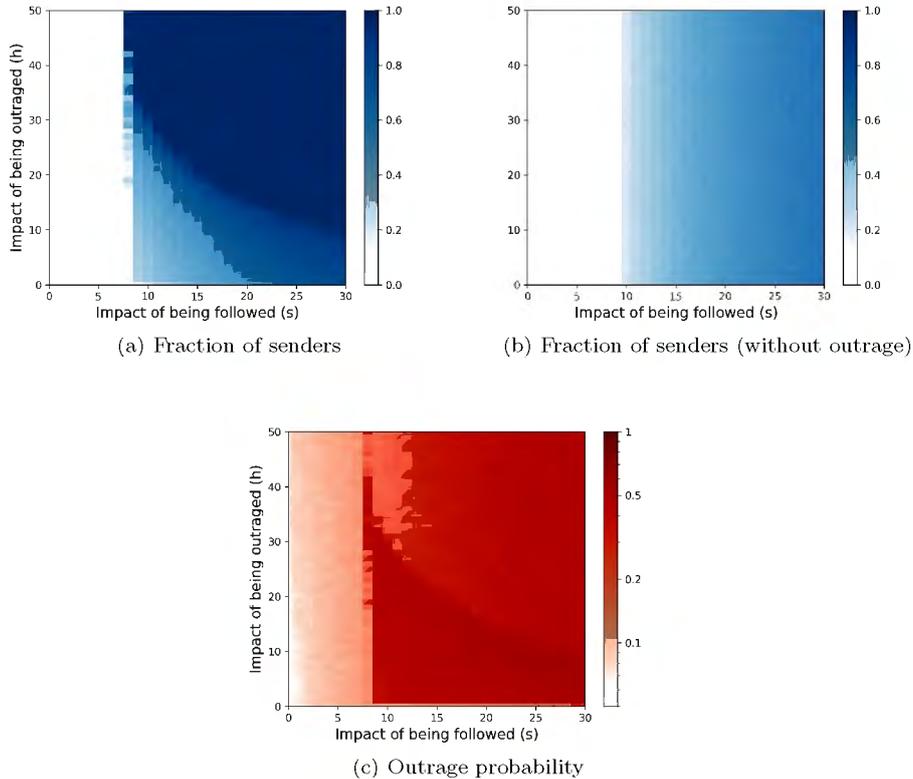

(a) Fraction of senders

(b) Fraction of senders (without outrage)

(c) Outrage probability

Figure 3: Results for one level of signaling, after a large number of rounds of simulation. **Top:** fraction of agents who invest in the binary display, as a function of the benefit of being followed $s$ and the cost of being outraged $h$, when agents can also invest in outrage, and (b) when they cannot. When agents can invest in outrage, signaling (blue regions) is obtained when the benefit of being followed is sufficiently large; near-uniform signaling (dark blue region) is obtained when the cost of being outraged is high. In the absence of outrage, at most 75% of individuals send. **Bottom:** outrage probability is maximal when signaling is non-uniform (light blue zone in (a)).
These simulations are computed with default values including: $h = 30$, $s = 10$, $p_1 = 0.1$, $c_1 = 30$, $c_2 = 5$; the population is composed of 200 agents; they can be affiliated with 2 other individuals and can receive up to 5 affiliation links (see Supplementary Information for more). Code and dynamic illustrations are available on this website.

We implement our model into an agent-based simulation. Contrary to the model, we consider a finite population, and implement local interactions. Individuals can choose a limited number of other agents to follow. The number of followers that an individual can have is also limited, in order to avoid winner-take-all effects.

Agents observe senders and non-senders directly, with probability $p_1$. In addition, they observe non-senders indirectly, through dyadic interactions with partners, who may express outrage at a non-sender they previously observed. Agents preferentially follow individuals whom they observe sending the signal (directly or indirectly), or whom they observe expressing outrage (during a dyadic encounter).

Agents interact based on two flexible behavioral traits: their investment in a binary display (one level of signaling), and their probability of expressing outrage at non-senders. In the initial round of simulation, these traits are set at 0. With a small probability, agents may try out another value of the trait.

Figure 3 shows attained fraction of senders in the case of a binary display, depending on the benefit of being followed $s$ and the cost of being outraged $h$. It also illustrates the crucial role of outrage in enabling uniform signaling.

## 5.2 Runaway costs

When signaling becomes uniform, onlookers can no longer determine who are the top-quality individuals. To attract followers, these individuals may find it in their interest to create and adopt a new discrete signal level, requiring an additional investment of $\Delta c_1(q)$. Again, we assume $\Delta c_1$ is a decreasing function of individual quality $q$. Over-performers have every incentive to advertise their increased investment — e.g. by finding new targets of outrage. We assume they may now pay $\Delta c_2$ to express outrage at individuals who are observed sending at the lower level, and guarantee visibility $p_3 > p_2$; targets lose $h$. Similarly to before, individuals are pushed to increase their investment in the signal (they are prevented from decreasing their investment to 0 for the same reasons as before). We expect full escalation to the new signal level when:

$$\Delta c_1(0) + \Delta c_2 < s + 2\sqrt{hs} \tag{5.1}$$

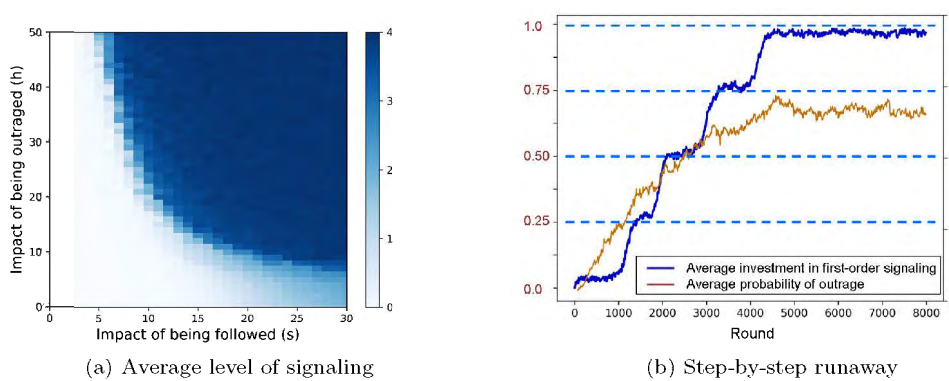

(a) Average level of signaling

(b) Step-by-step runaway

Figure 4: Results for four non-null, evenly spaced levels of signaling. **(a)** Average attained level of investment. Agents learn to invest in the highest signal level as soon as $h$ and $s$ are significant. **(b)** Step-by-step runaway, computed with a small value of $s = 2$ and a large value of $h = 150$, to show a clean ratchet effect. A dynamic illustration can be seen on the website.

Outrage could thus lead a population to adopt a costlier display. We relaunch our simulation with several evenly spaced levels of signaling (proportional costs). Agents may now express outrage at non-senders and lower-level senders (whom they still observe directly and indirectly). They preferentially follow: (i) first, an individual observed sending at level $n + 1$, (ii) second, an individual observed expressing outrage against a $n$-level sender, and (iii) third, a $n$-level sender.

When $h$ and $s$ are sufficiently large, outrage enables a step-by-step runaway process: individuals gradually learn to invest in the highest level of signaling (see Figure 4). This is in accordance with equation (5.1); when levels are evenly spaced, the marginal cost of signaling one level above is constant from one level to the next, and signal escalation may continue indefinitely. In reality, we expect marginal costs to increase at each step to infinity, as individuals are forced to miss out on increasingly important opportunities. The process will necessarily come to a halt. Eventually, high quality individuals will not benefit from creating a costlier display (and advertising it at the expense of others), and low quality individuals will prefer not to increase their investment, even if this means appearing relatively uncommitted.

## 6 Discussion

This paper offers a proof of concept for the existence of uniform, high-cost displays of commitment which serve to attract followers. The model is general, and may apply to other situations in which signalers compete for followers, and signaling seems exaggerated. Tentatively, our model could apply to high engagement on online social networks, and widespread prestige-motivated help in other species (Zahavi, 1995).

Our model is agnostic about any function the emerging behavior may serve at the level of the collective (e.g. encouraging group cohesion or cooperation; Atran & Henrich, 2010; Bulbulia & Sosis, 2011; Cimino, 2011; Durkheim, 1912/2008; Gambetta, 2009; Irons, 2001; Whitehouse & Lanman, 2014; Xygalatas et al., 2013). Uniform signals are explained at the individual level. Outrage benefits senders, by making their signal easier to spot. We show that, under certain conditions, outrage is sufficient to generate uniform signaling, and escalating costs.

We consider signals which take discrete values. Our model applies for binary displays of commitment, and for displays which categorize individuals (e.g. into participants of a high-ordeal ritual, of a low-ordeal ritual, and non-participants; Xygalatas et al., 2013). Of course this is a simplified vision of reality (Barker et al., 2019). Rituals do not occur in isolation, and receivers may make richer inferences by considering investment in related activities, or other qualities affecting a signaler's ability to invest in a display (e.g. status, Dumas et al., 2021). For instance, by broadening the temporal scope, we can look at long-term attendance in a frequent ritual, which is a continuous metric. Individuals who attend ritual activities more frequently are on average more generous towards other group members (Ruffle & Sosis, 2006; Soler, 2012), and are perceived as such (Power, 2017; Purzycki & Arakchaa, 2013).

Nevertheless, our focus on discrete rather than continuous signals should be seen as a feature of the model, and not a bug. Though continuously-valued signals are more informative, and appear more reasonable in a variety of situations, outrage requires clear-cut comparisons. In some cases, committed individuals could design discrete displays precisely for the purpose of expressing outrage.

We made the simplifying assumption that outrage is honest, in our model and simulation. Outrage is generally believed to be honest when hypocrites suffer sufficient retaliatory costs; yet retaliation against hypocrites is subject to much variation (Sommers & Jordan, 2022). In an extension of our model, we show that honesty can arise when hypocrites are preferential targets for others' outrage. Further research should investigate more systematically the conditions under which outrage is more likely to be honest, and/or treated as such by onlookers, ensuring that it can function as a second-order signal.

If we broaden the picture, second-order signaling can be seen as a specific case of signal amplification. The idea of amplifiers has been introduced to designate signals whose correlation with quality is indirect (Hasson, 1991). For instance, contour lines that accentuate margins on bird feathers or bars across feathers may have evolved as secondary features that make the primary signal, in this case healthy undamaged feathers, more conspicuous. Amplifiers may explain the sophistication of some mating signals. Contrary to signals, some amplifiers may not need to be costly to be reliable, as it is not in the interest of low-quality individuals to draw attention to their poor signaling (Gualla et al., 2008). In our model, outrage serves a function which is analogous to that of an amplifier: it increases the probability that the sender's signal will be detected. Outrage is a rather specific type of amplifier however, as it imposes costs on its target—through which uniform investment and runaway costs may emerge.

Our model may help explain mandatory displays of commitment, such as rites of passage (see also: Cimino, 2011; Densley, 2012; Gambetta, 2009; Iannaccone, 1992). Outrage can create a positive feedback loop, and sustain uniform, and therefore uninformative, displays. The resulting behavior is a specific type of norm. In general, norms can emerge from a variety of positive feedback loops, such as those created by social punishment or benchmark effects (Young, 2015). In our case, uniform displays arise endogenously, from the motivation to advertise one's prosocial commitment to group members, via first- and second-order signaling (we do not need to assume non-senders are punished).

Our model may also help explain exaggerated displays of commitment, e.g. during wartime (see also: Sosis et al., 2007; Whitehouse, 2018). Times of crisis tend to favor expression of commitment over others (Hahl et al., 2018), and may provide the initial push enabling signal runaway. In such cases, the system is expected to stop at extreme levels of signaling and outrage, pushing individuals to ever greater lengths to avoid appearing uncommitted. A similar logic may be at play with witch hunts or other collective crazes which follow a seemingly self-fulfilling pattern (Lotto, 1994).

The present model is kept minimal. It needs to be completed to explain why many uniform signals remain stable without reaching extreme values, or why, and when, they may deescalate. Depending on the context, individuals may look for commitment to other groups or values. Signals and non-signals can change meaning (e.g. pacifism instead of cowardice, or closed-mindedness instead of dedication to the group). We hope that our model can serve as a basis for investigation into these rich phenomena.

## Acknowledgements

We thank A. Sijilmassi for feedback on a early version of the manuscript, as well as two anonymous reviewers for their numerous and constructive comments. This research was supported by funding from the EURIP Graduate School for Interdisciplinary Research, and from the Agence Nationale pour la Recherche (ANR-17-EURE-0017, ANR-10-IDEX-0001-02).

# A  Supplementary data

A PDF containing a detailed description of the model and simulation, and their results can be found at https://doi.org/10.1016/j.jtbi.2023.111586.

The simulation code can be accessed online at https://evolife.telecom-paris.fr/outrage/.

Supplementary Information for

# Runaway signals:
Exaggerated displays of commitment may result from second-order signaling

# Contents

# S1 Baseline model

## S1.1 A multi-player model of costly signaling with uncertain observation

We adapt Gintis, Smith and Bowles' (2001) multi-player model, by making two small changes: importantly, we assume uncertain observation (with a certain probability $p_1 < 1$); more incidentally, we consider a continuous distribution of quality.

We consider a large group of individuals, who are characterized by a continuous quality $q$. We normalize minimum and maximum quality to 0 and 1 respectively: each individual's quality is drawn according to a continuous probability density function, whose support is $[0, 1]$. Individuals only observe their own quality. For mathematical convenience, the group is considered to be infinite in size.

Individuals alternate between two roles, that of signaler and receiver. Play occurs in three stages.

1. *Signaling stage.* Here, signalers decide whether to pay a cost $c_1(q)$ to send a signal, that is participate in a binary display of commitment. (The only other option is not to send.) Sending the signal is cheaper for high quality individuals: $c_1$ is a strictly decreasing continuous function of individual quality $q$, which takes positive values.

2. *Observation stage.* Here, receivers do two things. First, they decide whether to pay a small positive cost $\nu > 0$ to monitor the signal. Second, receivers who paid the cost of monitoring observe the action chosen by each individual signaler in the previous stage, i.e. whether the signaler opted to send or not send. The probability of observation is $p_1$ ($0 < p_1 < 1$). Since the population is infinite, they observe the behavior of a fraction $p_1$ of signalers. As long as sending occurs with positive probability, monitoring receivers each observe at least one sender (not necessarily the same one). Receivers who did not pay the cost of monitoring do not observe behavior in the signaling stage.

3. *Social interaction stage.* Here, receivers decide whether to follow one signaler, that is to affiliate to one individual from the group. Signalers gain positive payoff $s > 0$ for each receiver who decides to follow them. Receivers derive payoff $f(q')$ from following a signaler of quality $q'$, and null payoff from opting not to follow anyone. Following is on average beneficial, and high quality individuals are more desirable social partners: we assume $\mathbf{E}(f) > 0$, and that $f$ is a strictly increasing continuous function of the followee's quality $q'$. Following low quality individuals may or may not be detrimental (depending on the signal of $f(0)$).

Signalers may decide to send the signal depending on their quality. Receivers may decide to: not monitor and not follow anyone; not monitor and follow any individual, i.e. follow an individual chosen at random; monitor and follow a sender, i.e. follow an individual chosen at random among those signalers they observed sending the signal; or monitor and follow a non-sender. A pure strategy profile specifies: (i) when in the signaler role, whether to send or not send given own quality $q$, and (ii) when in the receiver role, which one of the above four strategies to play. We do not consider mixed strategies, in which individuals behave probabilistically.

## S1.2 Honest signaling equilibrium

### S1.2.1 Honest signaling strategy profile

There are two evolutionarily stable strategy profiles (ESS; Maynard Smith & Price, 1973). First, the strategy profile in which: (i) signalers never send, and (ii) receivers do not monitor, and follow an individual at random. This trivial strategy profile is always a strict Nash equilibrium, and therefore always an ESS. Indeed, deviation to sending the signal is costly for all signalers, whatever their quality. Deviation to monitoring is costly for receivers, as is deviation to not following—because following an individual at random is beneficial ($\mathbf{E}(f) > 0$). We ignore this ESS from here on.

The second ESS is obtained by considering a family of strategy profiles. For any threshold quality $\theta \in (0, 1)$, we define the honest signaling strategy profile for $\theta$ as the strategy profile whereby: (i) signalers send when their quality $q$ verifies $q > \theta$, and do not send when $q < \theta$, and (ii) receivers monitor, and follow a sender. We note this strategy profile $\mathrm{HS}(\theta)$. We do not consider signaler strategy given $q = \theta$; since quality is continuously distributed, this occurs with null probability. When individuals play according to such a strategy profile, we define $\pi(\theta) \equiv \mathbf{P}(q > \theta) \in (0, 1)$ the probability that a signaler is of relatively high quality $q > \theta$, and sends.

Any pure strategy equilibrium where signaling occurs with positive probability must follow this form. Indeed, note first that if receivers do not monitor the signal, signalers strictly lose from signaling, whatever their quality: signaling can only occur when senders positively affect their chances of being accepted, i.e. when receivers play according to (ii). Note second that $\theta$ must belong to $(0, 1)$:

if it is equal to 1, then signaling occurs with null probability; and if it is equal to 0, receivers strictly benefit from deviation to not monitoring.

The below demonstration further shows that signalers must play according to a threshold reaction norm of this form. We show that there can be only one honest signaling equilibrium, corresponding to a specific value of $\theta$, and second, that this equilibrium exists under a wide range of parameter values.

### S1.2.2 Characteristics of the honest signaling equilibrium

**Proposition 1** *HS($\theta$) is an ESS if and only if:*

$$\pi(\theta) = \frac{s}{c_1(\theta)} \tag{S1.1}$$

$$\nu < \mathbf{E}(f(q) \mid q > \theta) - \mathbf{E}(a) \tag{S1.2}$$

*Proof*: let us assume that individuals play according to the strategy profile HS($\theta$), for a given value of $\theta \in (0, 1)$. We first show that HS($\theta$) defines a strict Nash equilibrium if and only if both of the above conditions are verified.

HS($\theta$) is strict Nash if and only signalers of relatively high quality $q_H > \theta$, signalers of relatively low quality $q_L < \theta$, and receivers all stand to lose from deviation. We obtain equation [S1.1] by considering the case of signalers first. A signaler of quality $q$ can pay $c_1(q)$ to send, in which case she will face a fraction $p_1$ of well-disposed receivers in the future, who chose one individual to follow among the fraction $p_1 \times \pi(\theta)$ of the population that they observe sending the signal, their chosen followee earning $s$. Dividing the fraction of well-disposed receivers by the fraction of signals they chose from, we deduce that a sender on average recruits fraction $\frac{1}{\pi(\theta)}$ of receivers, and obtains an expected payoff of $-c_1(q) + \frac{s}{\pi(\theta)}$.

Signalers who do not send earn null payoff. By comparing the above expression to 0, we deduce that signalers of relatively high quality $q_H > \theta$ stand to lose from deviation iff $c_1(q_H) > \frac{s}{\pi(\theta)}$, and that signalers of relatively low quality $q_L < \theta$ stand to lose from deviation iff $c_1(q_L) < \frac{s}{\pi(\theta)}$. Since $c_1$ is a strictly decreasing function of quality, these two conditions are verified for all $q_H > \theta > q_L$ if and only if $c_1(\theta) = \frac{s}{\pi(\theta)}$; re-arranging, we obtain equation [S1.1]. (Note that signalers may send or not send indifferently when their quality $q$ is precisely equal to the threshold $\theta$; since this occurs with null probability, we neglect this possibility).

We obtain equation [S1.2] by considering next the case of receivers. A receiver pays $\nu$ to monitor the signal, and, since the pop-ulation is infinite, is certain to observe at least one signal, and ally with a signaler of relatively high quality; earning $\mathbf{E}(f(q) \mid q > \theta) - \nu$ on average. If she deviates to accepting at random, she gains instead $\mathbf{E}(f) > 0$; if she deviates to rejecting, she gains null payoff. By comparing these payoffs, we deduce that receivers can expect to lose from deviation if and only if condition [S1.2] is verified.

We have proven that HS($\theta$) is strict Nash if and only if conditions [S1.1-S1.2] are verified. Hence, under these conditions, the strategy profile is an ESS. Conversely, we show that when these conditions are not verified, HS($\theta$) is not an ESS: if $\theta$ is different to the critical quality determined by condition [S1.1], the previous reasoning shows that the strategy profile cannot be Nash, and therefore cannot be an ESS; and if the second condition [S1.2] is unverified, it can be invaded by a strategy profile in which receivers do not monitor and accept at random. This proves the proposed equivalence.

### S1.2.3 Existence of an honest signaling equilibrium

When satisfied, condition [S1.1] defines a unique critical quality $\theta$. Condition [S1.2] adds a constraint on $\theta$: the critical quality must be high enough to guarantee that the net gain from allying with a sender instead of an individual at random exceeds the cost of monitoring.

Gintis et al. show that when signaling is overly costly for low quality individuals ($c_1(0) > s$), an honest signaling equilibrium exists for a range of possibly binary distributions of quality. Below, we extend this result to continuous distributions of quality. When $c_1(0) > s$, an honest signaling equilibrium exists for a wide family of continuous distributions, depending on $\nu$ (those for which [S1.2] will be satisfied).

**Proposition 2** *When the signal is overly costly for the lowest quality signalers, there exists a range of possible values for the cost of monitoring $(0, \hat{\nu})$ for which an honest signaling equilibrium can be defined. In particular, there exists an honest signaling equilibrium where the cost of monitoring is arbitrarily small if and only if:*

$$c_1(0) > s \qquad (S1.3)$$

*Proof*: when $t$ varies in $[0, 1]$, $\pi(t)$ strictly decreases from 1, and $\frac{s}{c_1(t)}$ strictly increases from $\frac{s}{c_1(0)}$. Following the intermediate value theorem, a non-trivial critical quality $\theta \in (0, 1)$ which satisfies condition [S1.1] can be found if and only if condition [S1.3] is verified (Figure S1 gives a graphic argument). In addition, condition [S1.2] is verified if and only if the cost of monitoring is smaller than:

$$\hat{\nu} = \mathbf{E}(f(q)|q > \theta) - \mathbf{E}(f).$$

$\hat{\nu}$ is positive since $\theta$ is greater than the minimum quality 0. Condition [S1.2] is verified whenever the cost $\nu$ of monitoring is smaller than $\hat{\nu}$.

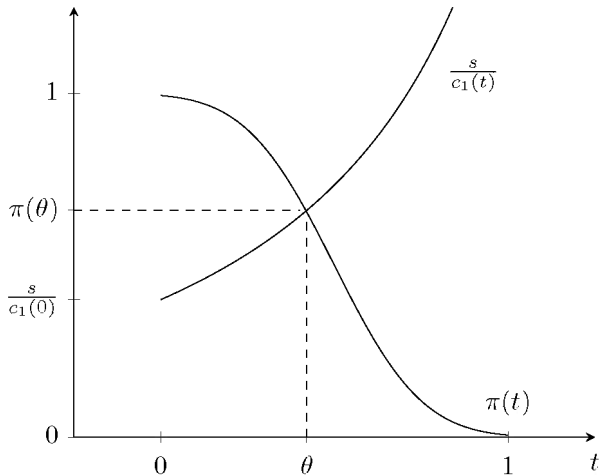

Fig. S1: Graphic determination of the critical threshold $\theta$

## S1.3 Interpretation

### S1.3.1 To evolve, a signal cannot be overly widespread

Following equation [S1.2], signaling can only be evolutionary stable when the relative benefit of conditioning alliance on the signal outweighs the cost of monitoring. In equilibrium, the signal is informative: when they observe the signal, receivers can infer the sender is of relatively high quality $q > \theta > 0$. More widespread signals (lower minimum bar $\theta$) are less informative to receivers, and less likely to evolve (depending on the cost of monitoring). In particular, a universal signal ($\theta = 0$) is always uninformative, and can never be evolutionarily stable (even when monitoring is free).

### S1.3.2 In equilibrium, desirable individuals signal and obtain a net benefit

Following equation [S1.1], the equilibrium value of the threshold quality $\theta$ is the value which balances cost $c_1(\theta)$ and benefit $\frac{s}{\pi(\theta)}$ of signaling. In equilibrium, desirable individuals of quality $q > \theta$ signal, and obtain a net benefit. When $\theta$ tends towards maximum quality 1, the benefit of signaling tends towards infinity: we can always expect signaling to emerge in the presence of a large motivated audience, since the first individuals to send will gain a large following.

When in contrast $\theta$ tends towards 0, the benefit of signaling falls to $s$. For signaling to remain informative, joining in with everyone else must be prohibitively costly for minimum quality individuals, i.e. we must have $c_1(0) > s$. Proposition 2 shows there is a form of equivalence; signals which are prohibitively costly for minimum quality individuals can evolve as long as monitoring is sufficiently cheap.

## S2 Runaway signal game

### S2.1 Adding outrage as a second-order signal

The signal runaway game occurs when we introduce outrage into the previous model. We view outrage as a *second-order signal*. Outrage refers to the commitment display (the first-order signal), by referring to a target's lack of signaling. Its function is to draw attention to the fact that the outraged individual did send the signal. Outraged senders increase everyone's incentive to send, and may destabilize the honest signaling equilibrium studied above. We modify the game in the following manner.

1. *Signaling stage.* Signalers decide whether to invest in costly signaling, as before, as well as whether to pay a cost $c_2 > 0$ to express outrage.

2. *Observation stage.* By expressing outrage, individuals draw attention to their signaling behavior. Signalers who paid the cost of second-order signaling $c_2$ are observed with increased probability $p_2 > p_1$ ($p_2 < 1$). Outrage is never observed in our model. Onlookers can only observe whether an individual sent the signal—with probability $p_1$ or $p_2$, depending on whether the individual paid to express outrage.

Outraged signalers observe the signal, and select a target. Each individual's signaling behavior is thus observed by receivers who pay the cost of monitoring, as well as signalers who pay the cost of outrage. We assume outraged signalers select a target among all individuals they observe not sending the signal. Since the population is infinite, outraged signalers find a target in all situations but uniform signaling.

3. *Social interaction stage.* Targets of outrage are harmed. Signalers lose $h > 0$ for each individual who expresses outrage against them. As before, receivers can follow signalers.

A pure strategy for the signaler now specifies whether or not to send, and, if opting to send, whether to express outrage or not, depending on own quality $q$. For every $q$, there are three possibilities: send and express outrage, send and do not express outrage, and do neither. Receiver strategies are unchanged.

Note that we do not consider the hypocritical strategy, whereby a signaler of quality $q$ does not send the signal, yet pays the cost of second-order signaling. Due to the simplified manner in which we model observation, this strategy is dominated. Hypocritical signalers pay $c_2$ to draw attention to their lack of signaling, which is never beneficial in our model because receivers do not observe outrage.

## S2.2 Effect of outrage on the previous signaling equilibrium

### S2.2.1 Honest signaling with outrage equilibrium

For any threshold quality $\theta \in (0, 1)$, we define the honest signaling with outrage strategy profile HSO($\theta$) as the strategy profile whereby: (i) signalers send and express outrage when their quality verifies $q > \theta$, and neither send nor express outrage when $q < \theta$, and (ii) receivers monitor, and follow a sender.

**Proposition 3** *HSO($\theta$) is an ESS if and only if:*

$$c_1(\theta) + c_2 = \frac{s}{\pi(\theta)} + \frac{\pi(\theta)h}{1 - \pi(\theta)} \quad (S2.1)$$

$$\nu < \mathbf{E}(f(q) \mid q > \theta) - \mathbf{E}(f) \quad (S2.2)$$

$$c_2 < \frac{p_2 - p_1}{p_2} \times \frac{s}{\pi(\theta)} \quad (S2.3)$$

Proof: analogous to the proof of Proposition 1. We assume that individuals play according to HSO($\theta$), for a given value of $\theta \in (0, 1)$. We show first that HSO($\theta$) defines a strict Nash equilibrium if and only if all three of the above conditions are verified.

Receiver strategy and trade-offs are unchanged, yielding condition (S2.2), which is identical to (S1.2). Each receiver observes a fraction $p_2 \times \pi(\theta)$ of senders, and chooses one to follow. An analogous calculation to the one before shows that a signaler of quality $q$ who sends and expresses outrage gains on average payoff: $-c_1(q) - c_2 + p_2 \times \frac{s}{p_2 \pi(\theta)} = -c_1(q) - c_2 + \frac{s}{\pi(\theta)}$.

Non-senders now face the cost of being potential targets of others' outrage. Each outraged signaler observes a fraction $p_1 \times (1 - \pi(\theta))$ of non-senders, and selects one as target. Non-senders face an outraged signaler with probability $\pi(\theta)$, and are observed by that individual with probability $p_1$. They can now expect a negative payoff, equal to: $p_1 \times \frac{\pi(\theta)(-h)}{p_1(1-\pi(\theta))} = -\frac{\pi(\theta)h}{1-\pi(\theta)}$. We obtain condition (S2.1) by comparing to the payoff above. When $\theta$ verifies this condition, signalers of quality $q = \theta$ are indifferent between sending both the signal and the second-order signal, and sending neither. Deviation to sending neither signal given $q > \theta$ is then detrimental, as is deviation to sending both signals given $q < \theta$.

Finally, condition (S2.3) is obtained by considering rare mutants who deviate to sending but not expressing outrage given $q > \theta$. Such an individual saves on the cost of second-order signaling $c_2$, but is observed with probability $p_1 < p_2$, earning only $p_1 \times \frac{s}{p_2 \pi(\theta)}$ on average in terms of followers. Comparing to the payoff of an outraged sender, we obtain the proposed condition.

To conclude, we have proven that HSO($\theta$) is strict Nash if and only if conditions [S2.1-S2.3] are verified. Under these conditions, the strategy profile is an ESS. Conversely, we can show that both of the first two conditions are necessary, using an analogous argument to the one in Proposition 1. In addition, the last condition is necessary because otherwise rare mutants who deviate to sending but not expressing outrage given $q > \theta$ could invade. This proves the proposed equivalence.

### S2.2.2 Sufficient condition for the evolution of outrage when signaling is honest

**Proposition 4** *When receivers follow conditionally on the signal, senders all invest in outrage if:*

$$c_2 < \frac{p_2 - p_1}{p_2} s \tag{S2.4}$$

Proof: since $\frac{1}{\pi(\theta)} \geq 1$, the above constitutes a sufficient condition for (S2.3), that is a sufficient conditions for senders to lose from deviation to not expressing outrage when individuals play according to HSO($\theta$), for a certain $\theta \in (0,1)$.

The only other family of pure strategies in which receivers follow conditionally on the signal are strategies that are akin to the baseline honest signaling equilibrium, whereby senders opt not to express outrage. In such a situation, there exists $\theta \in (0,1)$ such that senders are observed with probability $p_1$, and gain $\frac{s}{p_1 \pi(\theta)}$ when observed. Deviation to expressing outrage costs $c_2$ and increases one's visibility, leading to relative benefit $(p_2 - p_1)\frac{s}{p_1 \pi(\theta)} > (p_2 - p_1)\frac{s}{p_2}$. When the above condition holds, that deviation is net beneficial, and outrage invades.

### S2.2.3  Condition under which $HSO(\theta)$ cannot be an ESS

Condition [S2.1] captures the effect of outrage on the equilibrium value of $\theta$. When $c_2 < \frac{\pi(\theta)h}{1-\pi(\theta)}$, we obtain a lower threshold than in the baseline case. Outrage then increases the incentive to signal, pushing more individuals to send both signals. Under certain conditions, the minimum bar $\theta$ will be pushed all the way to 0. When this occurs, honest signaling can no longer be stable. The below proposition gives a sufficient condition for this to happen.

**Proposition 5** *For every positive threshold $\theta$, HSO($\theta$) is not an ESS if:*

$$c_1(0) + c_2 < s + 2\sqrt{hs} \tag{S2.5}$$

*Proof*: For HSO($\theta$) to be an equilibrium, $\pi_S = \pi(\theta)$ must verify equation [S2.1]. Multiplying by $\pi_S(1 - \pi_S)$ ($\pi_S$ is always positive and smaller than 1 at such an equilibrium), we obtain equivalently:

$$(c_1(\theta) + c_2 + h)\pi_S^2 - (c_1(\theta) + c_2 + s)\pi_S + s = 0$$

We recognize a second-order equation in $\pi_S$, whose discriminant is equal to:

$$\Delta = (c_1(\theta) + c_2 + s)^2 - 4(c_1(\theta) + c_2 + h)s = (c_1(\theta) + c_2 - s)^2 - 4sh$$

Outrage will push $\theta$ all the way to 0 when the above equation has no solution in the interval $(0,1)$. A sufficient condition for that to occur is $\Delta < 0$. Since $c_1(\theta)$ increases when $\theta$ decreases, and since we necessarily have $c_1(0) + c_2 > c_1(0) > s$ (otherwise there is no signaling equilibrium to start from following Proposition 2), we deduce that the squared term is positive when $\theta$ is sufficiently small. We can then take the squared root and obtain a sufficient condition by replacing $\theta$ with 0; we obtain the proposed condition.

## S2.3 Uniform signaling can be stable when outrage harms ambiguous targets

### S2.3.1 Extension to ambiguous secondary targets of outrage

Our main result is therefore negative: if outrage is sufficiently cheap to express, as per condition (S2.4), and being the target of others' outrage is sufficiently costly, as per condition (S2.5), then outrage invades, and fully destabilizes any honest signaling equilibrium. Under such conditions, there can be no signaling ESS. Uniform signaling remains impossible here, because the function of outrage is merely to attract more followers, and receivers stop monitoring the signal when it is uniform.

Uniform signaling can however be made possible by extending the target selection mechanism. When all individuals signal, there are no non-senders to target. In our model, for technical reasons, this does not prevent signalers from investing in second-order signaling (because the model occurs in separate stages for simplicity, and we need outraged signalers' visibility to increase before the observation stage). We may instead assume that when individuals do not find non-senders, they use more ambiguous targets instead, in order to express outrage.

We modify our model, by having outraged senders select as target: (1) a non-sender whom they observe, or, if they do not observe any non-sender (because signaling is uniform) (2) a signaler whose behavior they do not observe. Second-order signaling now serves two functions. When senders aren't observed, they miss out on possible followers *and* risks being the target of others' outrage.

### S2.3.2 Uniform signaling with outrage equilibrium

Let us consider the universal signaling with outrage (USO) strategy profile, whereby: (i) signalers send and express outrage whatever their quality, and (ii) receivers do not monitor the signal, and accept a signaler at random.

**Proposition 6** *USO is an ESS if and only if:*

$$c_2 < (p_2 - p_1) \times \frac{h}{1 - p_2} \qquad (S2.6)$$

*Proof*: let us assume individuals play according to the USO strategy profile. Since receivers do not monitor the signal, senders do not recruit more followers than non-senders. All signalers send and express outrage, by targeting one of the $1 - p_2$ individuals they each do not observe sending. With probability $1 - p_2$, a signaler will constitute a potential (ambiguous) target for another signaler; dividing, we deduce that each individual loses $h$, on average.

No individual benefits from deviation to not sending. Any individual who does so risks become a priority target for other individuals with probability $p_1$, and faces an infinite loss. If an individual opts not to express outrage, she saves on cost $c_2$, but increases her chance of constituting a target for others from $1 - p_2$ to $1 - p_1$, losing $\frac{1-p_1}{1-p_2}h$ on average. By comparing with $h$, we deduce that USO is strict Nash, and therefore ESS, if (S2.6) holds. Conversely, if this condition is unverified, senders do not lose from deviation to not expressing outrage; mutants who do not express outrage can then invade. This proves the proposed equivalency.

### S2.3.3 Sufficient condition for outrage

Under the conditions derived in this section, outrage may transform the honest signaling equilibrium into a stable equilibrium where all individuals signal, and the signal is completely uninformative. When condition (S2.5) is verified, outrage should push all individuals to signal, destabilizing the honest signaling strategy profile. As long as it is sufficiently cheap, as per condition (S2.6), we may end up with generalized signaling.

More precisely, we derive a sufficient condition for outrage to exist in all the potential situations under consideration. To simplify, we assume $\nu = 0$ in the below proposition; such that we should either be in a case of the form HSO($\theta$), when $\theta > 0$, and otherwise be in the case of USO.

**Proposition 7** *When monitoring is free ($\nu = 0$), in any ESS where signaling occurs with positive probability, senders express outrage if:*

$$c_2 < (p_2 - p_1) \times \min\{\frac{s}{p_2}, \frac{h}{1-p_2}\} \qquad (S2.7)$$

*Proof*: let us assume we are in an ESS where signaling occurs with positive probability, and where senders express outrage. Since the cost of sending both signals $c_1(q) + c_2$ is a decreasing function of individual quality $q$, signaler behavior can be described according to a threshold $\theta \in [0, 1)$ above which they send both signals.

If $\theta > 0$, we must be in the case of honest signaling with outrage. Since $\nu = 0$, receivers strictly benefit from using the signal. Let us consider a signaler of quality $q \geq \theta$, who sends both signals, and earns on average $p_2 \times \frac{s}{p_2 \pi(\theta)} - c_1(q) - c_2$. Were such an individual to deviate to not expressing outrage, she would save on the cost of outrage $c_2$, but decrease her chances of being observed from $p_2$ to $p_1$. On average deviation to not expressing outrage for a sender leads to payoff differential: $c_2 - (p_2 - p_1)\frac{s}{p_2\pi(\theta)} \leq c_2 - (p_2 - p_1)\frac{s}{p_2}$. Since we are in an ESS, and since $\theta < 1$, we deduce that we must have: $c_2 < (p_2 - p_1)\frac{s}{p_2}$.

If $\theta = 0$, we must be in the case of the USO ESS, and therefore have $c_2 < (p_2 - p_1)\frac{h}{1-p_2}$, following Proposition 6. This proves the implication.

Finally, let us assume instead that players are playing according to a strategy profile in which signaling occurs with positive probability, and senders do not express outrage. We prove the strategy profile cannot be ESS when the above condition holds. To do this, note first that we must be in (a situation akin to) the baseline honest signaling equilibrium. The same steps as in the proof of Proposition 4 show that deviation to expressing outrage is net beneficial under the above condition. The strategy profile under consideration can therefore not be an ESS. This proves the proposed equivalency.

# S3 Simulation

## S3.1 Presentation of the simulation

The multi-agent simulation, written in Python, is based on the *Evolife*[1] platform. Agents differ by their quality. Agent qualities are uniformly distributed between 0 and 100. They may signal at a certain level at a cost that smoothly decreases with their quality. Agents learn two features through a simple local search: their investment in signaling and their probability of expressing outrage (investment in signal monitoring is an optional learned feature). A typical example of run can be see on the website.

All interactions in the simulation are meant to be local. The population is structured in groups. Individuals meet each other in a randomized order within groups. During their first encounter (Algorithm 1), they observe each other's signal with a certain probability which depends on a global parameter called *InitialVisibility* (parameter $p_1$ in the model and on a feature, *MonitoringProbability* ($\nu$ in the model) (set to 1 by default, but that can be learned by individuals as an option).

---

[1] All programs are open source and are available at this Website: https://evolife.telecom-paris.fr/outrage. The program described here can be found in the *Evolife* package at `Evolife/Apps/Patriot/Patriot.py`

**Algorithm 1** Observe

**Input:** $self, Partner$
**if** $random() \leq self.MonitoringProbability$
**and** $random() \leq InitialVisibility$ **then**
    **if** $self.signal < Partner.signal$ **then**
        ▷▷ *self remembered as potential outrage target*
        **add** $(self, self.signal)$ **to** $Partner$'s **outrage memory**
    **end if**
    ▷▷ *self remembered as potential affiliation target*
    **add** $(self, self.signal)$ **to** $Partner$'s **affiliation memory**
**end if**

During a second randomized encounter (Algorithm 2), individuals may express outrage toward third parties. The point of outrage is to indicate that one's own signal is superior to the target's signal (this translates in the apparent signal $Target.signal + 1$ in the algorithm). Each individual learns a feature named *OutrageProbability* and decides to be outraged accordingly.

**Algorithm 2** Outrage

**Input:** $self, Partner$
**if** $random() \leq self.OutrageProbability$ **then**
    ▷▷ *self communicates outrage target*
    $Target \leftarrow$ **worst individual in self's outrage memory**
    **if** $Target.signal < Partner.signal$ **then**
        **add** $(Target, Target.signal)$ **to** $Partner$'s **outrage memory**
    **end if**
    **add** $(self, Target.signal + 1)$ **to** $Partner$'s **affiliation memory**
**end if**

In a third randomized encounter, individuals attempt to establish friendship based on the observed signals (Algorithm 3).

**Algorithm 3** Interact

**Input:** $self, Partner$
**if** $Partner$ in $self$'s **affiliation memory then**
    $PartnerSignal \leftarrow Partner$'s **memorized signal**
**else**
    $PartnerSignal \leftarrow 0$
**end if**
**if**   self's **affiliation set is not full**
**or**  $PartnerSignal \geq self$'s **current worst friend's signal then**
    ▷▷ *Partner becomes self's friend*
    $self.affiliate(Partner, PartnerSignal)$
**end if**

After these three rounds, payoffs are computed (Algorithm 4):

individuals get rewarded for having attracted affiliates (they receive $FollowerImpact$, corresponding to parameter $s$ in the model) and for being affiliated with high quality individuals (they receive $FollowingImpact \times Partner.Quality$ for each partner; function $a(q')$ in the model). Individuals get punished if they were the target of outrage (parameter $h$ in the model). Agents' memory is reset after the assessment phase. However, they store payoffs and learn periodically from them. Agents have a limited lifespan and get fully reinitialized when being reborn with the same quality.

---

**Algorithm 4** Assessment

**Input:** $self$
**for** $F$ in $self$'s friends **do**
    ▷▷ *payoff for having attracted a follower (s)*
    $F.Points\ +\!\leftarrow FollowerImpact$
    ▷▷ *payoff for being affiliated with F (depends on F's quality)*
    $self.Points\ +\!\leftarrow FollowingImpact \times F.Quality$
**end for**
$self.Points\ -\!\leftarrow$ **cost of signaling**
$self.Points\ -\!\leftarrow OutrageCost \times self.OutrageProbability$
$self.Points\ -\!\leftarrow MonitoringCost$
**if** $self.Outrage$ **then**
    $Target \leftarrow self$'s **outrage memory worst individual**
    ▷▷ *outrage target is harmed*
    $Target.Points\ -\!\leftarrow OutragePenalty$
**end if**
$self.resetMemory()$

---

## S3.2 Differences between model and simulation

In the model, we consider an infinite population, such that one individual's strategy does not affect overall probabilities. In addition, receivers may monitor, observe and choose senders in a perfectly balanced way. In contrast, the simulation program is meant to implement a more realistic setting in which all interactions remain local. The population is periodically split into random groups within which interactions occur. Agents meet systematically, though in a randomized order. An agent may or may not see the partner's signal, and may or may not express outrage at some previously seen individual, in order to prove its own signaling to the partner. Due to locality and chance, there is a variance in the number of affiliates each visible sender may attract. To prevent a winner-take-all effect, we limit the number of affiliations per individual and the number of affiliates each sender may recruit.

Another divergence with the model lies in the payoff function $f(q)$ that depends on the quality of the individual with whom one gets affiliated. The role of this function in the model is to motivate agents to search for high-quality individuals to affiliate with. In the simulation, we made a simplifying assumption and hard-wire the preference for intense signals.

Another difference comes from the fact that agents do not always adopt the ideal strategy corresponding to their quality. They need time to learn their various options (sending the signal, expressing outrage) and they constantly explore alternatives with a certain probability. Despite behavioral variance due to chance and to this "learning noise", the simulation is robust, i.e. it produces similar outcomes for a wide range of parameter values.

Variance can be seen as an advantageous feature of the simulation. When all individuals end up sending the same signal, there are no obvious outrage targets. Hence the possibility introduced in the model of expressing outrage at ambiguous individuals, i.e. individuals that either do not send or were not observed while sending. By contrast, in the simulation, the constant existence of exploring individuals maintains potential outrage targets.

### S3.3 Parameters

The simulation program relies on a variety of parameters. The most relevant ones are listed in table 1. Individuals get 'Follower Impact' ($s$ in the model) for each agent that affiliates with them. The 'Signaling cost coefficient' provides the scale of signal cost: it corresponds to the the cost $c_1$ paid by a medium-quality Sender. 'Signaling cost decrease' controls the variation of signaling cost depending on quality ($c_1(q)$ in the model) (0: no variation; 1: linear decrease; higher values: steeper, non-linear convex decrease). 'Outrage penalty' ($h$ in the model) is endured by individuals each time they are someone's outrage target. The parameter 'Outrage cost' implements a gradual version of model's fixed cost $c_2$: outraged individuals pay a cost that is proportional to this parameter and to their (learned) propensity to express outrage. 'Initial visibility' is the probability of individuals' signals to be seen during the observation round ($p_1$ in the model). Finally, two parameters control the learning speed. For each learned feature, the value explored next may totally change according to 'Jump probability' or locally change according to 'Additive exploration'.

Parameters' values are systematically explored in the simulations of the next section, while non-varying parameters are set to the default values of table 1.

| Description | Default value |
|---|---|
| Population size | 200 |
| Number of groups | 5 |
| Maximum number of followers (affiliates) | 5 |
| Number of affiliations (followees) | 2 |
| Impact of being followed ($s$) | 10 |
| Impact of outrage ($h$) | 30 |
| Signaling cost coefficient | 100 |
| Signaling cost decrease | 5 |
| Outrage cost ($c_2$) | 5 |
| Initial visibility ($p_1$) | 0.1 |
| Jump probability | 0.05 |
| Additive exploration | 20% |

Table 1: List of most relevant parameters.

## S3.4 One signal level

The emergence of uniform signaling due to outrage is a robust phenomenon that occurs for a wide range of parameters (see figures in the main article and dynamic examples on the website.

Figure S2a shows that three regions can be distinguished, based on costs and payoffs: a no-signal zone, a uniform signaling zone (dark blue) and an intermediary zone (light blue) corresponding to a separating equilibrium with a smaller fraction of senders. Uniform signaling (dark blue region) is obtained for low values of $c_1$ and high values of $s$. Figure S2b reveals that outrage is maximally probable in the intermediary zone, where it is used by agents as a way to increase the probability of being perceived as sender.

Figure S3 shows investment in both first- and second-order signaling, depending on signaling costs. The figure reveals the role of outrage as signal enhancer: in Fig. S3a, uniform signaling (dark blue region) expands toward costly signals at the bottom where outrage is cheap; in Fig. S3b we can see that outrage is intense in the separating equilibrium zone that corresponds to the light blue zone in Fig. S3a.

Figure S4a shows that uniform signaling (dark blue) emerges when visibility ($p_1$) is low and outrage cost ($c_2$) is not too high. For other values of visibility, the separating equilibrium (light blue) is observed except when outrage is free. Figure S4b clearly shows that outrage promotes uniform signaling: outrage probability is sig-

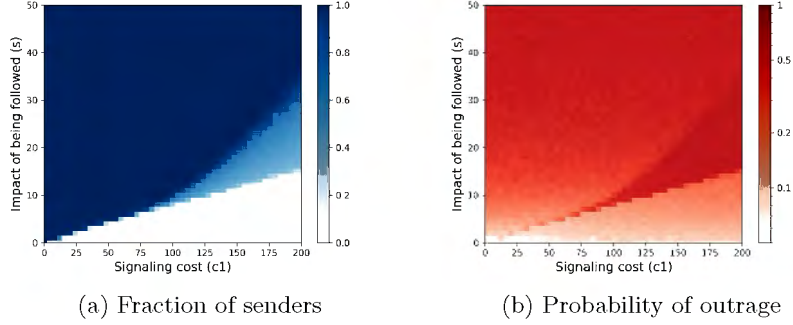

(a) Fraction of senders  (b) Probability of outrage

Fig. S2: First- and second-order signal after many rounds, depending on signaling cost $c_1$ and follower impact $s$. (a) Fraction of senders; (b) average probability of outrage.

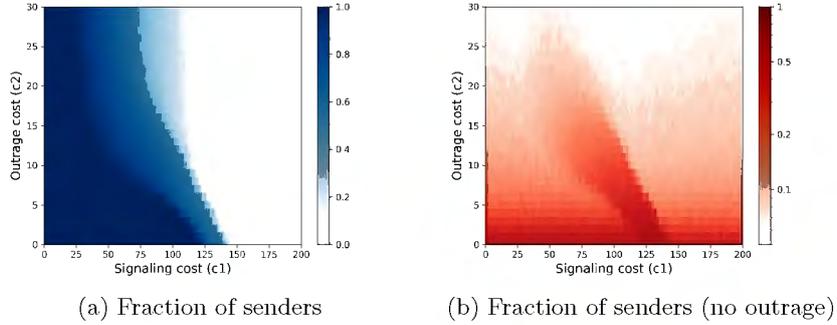

(a) Fraction of senders  (b) Fraction of senders (no outrage)

Fig. S3: First- and second-order signal after many rounds, depending on the signaling cost coefficient $c_1$ and the cost of expressing outrage $c_2$. (a) Fraction of senders; (b) average probability of outrage.

nificant in the zone that corresponds to uniform signaling and where outrage is cheap.

Figure S5 shows that the emergence of a signaling situation depends on two learning parameters. The first one controls the agents' maximal additive exploration during the learning of features (here signal and outrage probability). The jump probability coefficient controls the probability of "jumping" to any value from time to time. A moderate value of either parameter is necessary for learning to function properly. Too large values generate mere noise.

Figure S6a shows the necessity of an imbalance in the number of affiliations and the number of affiliates per individual. For uniform signal to emerge in the simulation, the benefit of attracting $k$ affiliates beyond the expected value (i.e. beyond the number of affiliations) must exceed the cost of enduring outrage (here $k \times 10 \geq 30$). Note that when individuals can have only one affiliation, the top half of them become senders and attract all available votes (hence the light-blue vertical line in figure S6a).

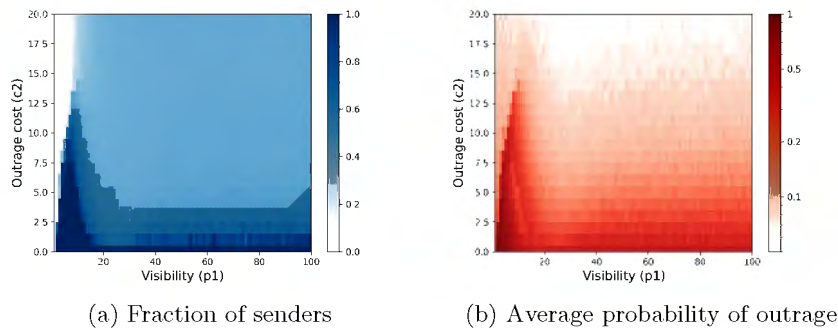

(a) Fraction of senders  (b) Average probability of outrage

Fig. S4: Fraction of senders and average probability of outrage, as a function of visibility $p_1$ and outrage cost $c_2$.

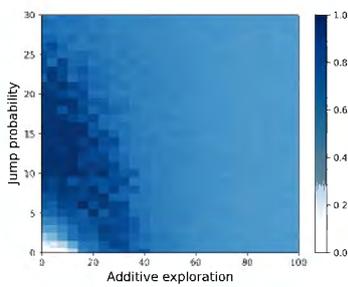

Fig. S5: Fraction of senders (blue shades) depending on additive exploration and jump probability.

The population in the simulation is finite. It is structured in randomly drawn groups in which interactions occur (groups are periodically redrawn). Figure S6b shows the proportion of signalers as a function of the number of groups and the size of the population. We can observe that groups should be neither too small nor too large for signaling to emerge. In a very small group, all individuals attract the maximum number of affiliations anyway and sending the signal is useless (white region). In a large group, enough individuals are visible to each agent (up to the numnber of affiliates it can accept) and outrage becomes useless (light blue region).

In addition, Figure S7 shows how attained investment in signaling varies with the 'Signaling cost coefficient' (variation of $c_1(q)$ in the model). It reveals that cost inequality between the low-quality (or least motivated) individuals and the high-quality (or highly motivated) individuals promotes runaway toward high-cost signal levels.

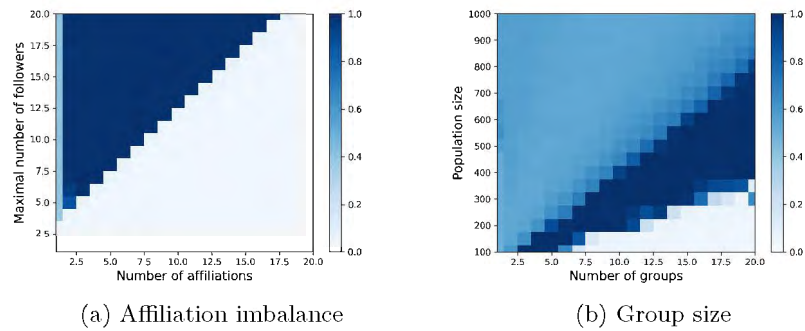

(a) Affiliation imbalance  (b) Group size

Fig. S6: Fraction of senders as a function of (a) the number of given and received affiliation links and (b) the number of groups vs. the population size.

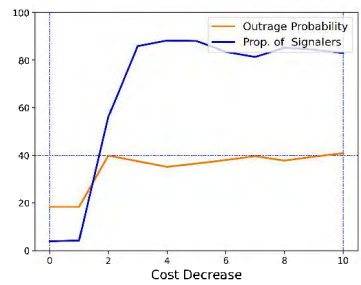

Fig. S7: Average level of signaling as a function of the 'Signaling cost decrease' coefficient.